\documentclass{elsart}

\newif\ifpdf
	\ifx\pdfoutput\undefined
	\pdffalse    
\else
	\pdfoutput=1 
	\pdftrue
\fi
\ifpdf
	\usepackage[pdftex]{graphicx}
\else
	\usepackage{graphicx}
\fi
\usepackage{amsfonts}

\begin{document}

\ifpdf
	\DeclareGraphicsExtensions{.pdf}
\else
	\DeclareGraphicsExtensions{.eps}
\fi

\begin{frontmatter}
\title{Chern--Simons Hadronic Bag from Quenched Large-$N$ QCD}\thanks{Accepted
for publication on Phys. Lett. B}

\author[Trieste]{Stefano Ansoldi}
\thanks{e-mail address: ansoldi@trieste.infn.it},
\author[Atlanta]{Carlos Castro}
\thanks{e-mail address: castro@ctsps.cau.edu},
\author[Trieste]{Euro Spallucci}
\thanks{e-mail address: spallucci@trieste.infn.it}
\address[Trieste]{Dipartimento di Fisica Teorica,
				  Universit\`a di Trieste
		          and INFN, Sezione di Trieste}
\address[Atlanta]{Center for Theoretical Studies of Physical Systems,
				  Clark Atlanta University, Atlanta, GA.30314 }
\begin{abstract}
$\mathrm{SU}(N)$ reduced, quenched, gauge theories
have been shown to be related
to string theories. We extend this result and
show how a $4$-dimensional,
reduced, quenched, Yang--Mills theory, supplemented by the
topological term, can be related through the
Wigner--Weyl--Moyal correspondence to an open
\textit{$3$-brane\/} model.
The boundary of the $3$-brane is described by a
Chern--Simons $2$-brane. We identify the bulk of the $3$-brane with the
interior of a hadronic bag and the world-volume of the
Chern--Simons $2$-brane with the
dynamical boundary of the bag. We estimate the value of the induced bag
constant  to be a little less than $200 \, \mathrm{MeV}$.
\end{abstract}
\end{frontmatter}

Recent proposals for a  non-perturbative formulation of String Theory
\cite{ikkt} have renewed the interest for matrix models of non-Abelian
gauge theories.
Large-$N$ Yang--Mills theories on a $D$-dimensional spacetime
\cite{largen}
have been shown to be equivalent to reduced matrix models, where the
original $N \times N$ matrix gauge field
${\mbox{\boldmath{$A$}}}_\mu {}^i_j\left(
x\right)$ is replaced by the same
field at a single point, say $x^\mu=0$
\cite{reduced} (for a recent review see \cite{makee}).
Partial derivative operators
are replaced by commutators with a fixed diagonal matrix
${\mbox{\boldmath{$p$}}}_\mu {}^i_j$, playing the role of translation
generator and called the \textit{quenched momentum\/} \cite{quench}.
Accordingly, the covariant derivative becomes
$\mathrm{i} {\mbox{\boldmath{$D$}}}_\mu =
\left[ {\mbox{\boldmath{$p$}}}_\mu
+ {\mbox{\boldmath{$A$}}}_\mu ,\dots
\right]$. Thus, the reduced, quenched,
Yang--Mills field  strength is
\[
{\mbox{\boldmath{$F$}}}_{\mu\nu}{}^i_j\equiv
\left[\mathrm{i} {\mbox{\boldmath{$D$}}}_\mu,
\mathrm{i} {\mbox{\boldmath{$D$}}}_\nu
\right]^i_j,
\]
which leads to the \textit{matrix\/} Yang--Mills
action in four dimensions
\begin{equation}
S_{\mathrm{qYM}}^{\mathrm{red.}}=-{1\over 4}
\left( {2\pi\over a}\right)^4
{N\over g^2_{\mathrm{YM}}}  \mathrm{Tr}{\mbox{\boldmath{$F$}}}_{\mu\nu}
{\mbox{\boldmath{$F$}}}^{\mu\nu},
\label{redym}
\end{equation}
where $g_{\mathrm{YM}}$ is the strong
coupling constant and $a\equiv 2\pi/
\Lambda$ is an inverse momentum  cut-off, or lattice spacing.
$S_{\mathrm{qYM}}^{\mathrm{red.}}$
is the starting point to study large-$N$
Yang--Mills theory\footnote{The subscript ``${}_{\mathrm{q}}$'' means
\textit{quenched}, not to be confused with ``quantum''.}.
For our purposes it is important to supplement
$S_{\mathrm{qYM}}^{\mathrm{red.}}$ with the
contribution from topologically
non-trivial field configurations, i.e. instantons, which is accounted
for by the \textit{topological term\/} $\epsilon^{\lambda\mu\nu\rho}
F_{\lambda\mu} F_{\nu\rho}$ (the reason for this
choice will be evident later on). Thus
\[
S_{\mathrm{qYM}}^{\mathrm{red.}}\longrightarrow
S_{\mathrm{q}}^{\mathrm{red.}}\equiv S_{\mathrm{qYM}}^{\mathrm{red.}}
+S_{\mathrm{q}\theta}^{\mathrm{red.}},
\]
where
\[
S_{\mathrm{q}\theta}^{\mathrm{red.}}= -{\theta N g^2_{\mathrm{YM}}
\over 16\pi^2}
\left( {2\pi\over a}\right)^4 \epsilon^{\mu\nu\rho\sigma}
 \mathrm{Tr}  {\mbox{\boldmath{$F$}}}_{\mu\nu}
{\mbox{\boldmath{$F$}}}_{\rho\sigma}
\]
and $\theta$ is the vacuum angle. \\
The matrix model (\ref{redym}) in the limit $N\to\infty$ has been shown
to describe strings \cite{bars}. Here we would like to extend this
result and show that the model encoded by
$S_{\mathrm{q}}^{\mathrm{red.}}$
includes, in the large-$N$
limit, not only strings, but also an open
$3$-brane with a Chern--Simons
$2$-brane as its dynamical boundary.
These higher dimensional objects would provide
appropriate  models for hadronic bags embedded in a
$4$-dimensional target
spacetime. In this way we would hopefully  fill
the gap between QCD, as the fundamental quantum field theory of strong
interactions,  and more phenomenologically oriented models for strongly
interacting, confined objects. To make evident the relationship between
matrix gauge fields and extended objects we
shall implement the effective
and simple correspondence among unitary operators
in Hilbert space and ordinary functions in a
non-commutative phase space
\cite{madore}, which has been originally established by the
Wigner--Weyl--Moyal (WWM) formulation of
quantum mechanics (see \cite{mww}
for recent applications of this quantization method).\\
The eigenvalues of the unitary operator
${\mbox{\boldmath{$D$}}}_{\mu} {}^i_j$
span a $D$-dimensional eigen-lattice
\cite{sochi} with $ 0 \le D \le 4$.
Thus ${\mbox{\boldmath{$D$}}}_\mu $ can be expressed in terms of $2D$
independent matrices, ${\mbox{\boldmath{$p$}}} _{i}$,
${\mbox{\boldmath{$q$}}}_j$,
$i, j = 1,\dots, D$, through the WWM relation
\begin{equation}
{\mbox{\boldmath{$D$}}}_\mu \equiv {1\over (2\pi)^D}\int \d^D p \d^D q
{\mathcal{A}}_\mu( k, z)
\exp\left( i q^i  {\mbox{\boldmath{$p$}}}_i +  i p^j
{\mbox{\boldmath{$q$}}}_j\right) ,
\label{wwm}
\end{equation}
where the operators ${\mbox{\boldmath{$p$}}}_i$,
${\mbox{\boldmath{$q$}}}_j$
satisfy the Heisenberg algebra
\[
\left[ {\mbox{\boldmath{$p$}}}_i, {\mbox{\boldmath{$q$}}}_j\right]=
-i \hbar \delta_{ij}
\]
and $\left( q^i, p^j\right)$ play the role of coordinates in a
Fourier dual space. $\hbar$ is the \textit{deformation parameter},
which for historical
reason is often represented by the same symbol
as the Planck constant.\\
The basic idea under this approach is to identify the Fourier
space as the dual of a $(D+D)$-dimensional
world-manifold of a $p=2D-1$ brane.
The case $D=1$, corresponding to strings,
has been investigated in depth because
of the relation between the $\mathrm{su}(\infty)$ Lie algebra and
$\mathrm{sdiff}\left(\Sigma\right)$, the area
preserving diffeomorphism  algebra over a $2$-dimensional
manifold $\Sigma$. Much less
attention has been given, in this framework,
to higher dimensional objects
(a remarkable exception is \cite{gaba}).\\
In this letter we shall consider the case $D=2$ and
show that in the limit $N\to\infty$ the action
$S_{\mathrm{q}}^{\mathrm{red.}}$
becomes a  bag action endowed with a dynamical boundary.

The WWM relation establishes a one-to-one
correspondence between a linear
operator, $ {\mbox{\boldmath{$D$}}}_\mu $ in our case,
acting over a Hilbert
space ${\mathcal{H}}$ of square integrable functions
on ${{\Rset}}^{D}$ and a
smooth function ${\mathcal{A}}_\mu( x, y)$, which is the anti-Fourier
transform of  ${\mathcal{A}}_\mu( k, z)$ in (\ref{wwm}):
\begin{eqnarray}
&& {\mathcal{A}}_\mu( q, p)={1\over N}\mathrm{Tr}_{{\mathcal{H}}}
\left[ {\mbox{\boldmath{$D$}}}_\mu
\exp\left( -\mathrm{i} {\mbox{\boldmath{$p$}}}_i q^i -\mathrm{i}
{\mbox{\boldmath{$q$}}}_j p^j\right)\right]
\nonumber \\
&& {\mathcal{A}}_\mu( x, y)=\int \d^Dq  \d^Dp
{\mathcal{A}}_\mu( q, p)
\exp\left(  \mathrm{i} q_i x^i +\mathrm{i} p_j y^j\right),
\nonumber
\end{eqnarray}
where $\mathrm{Tr}_{{\mathcal{H}}}$ means the
sum over diagonal elements with respect
to an orthonormal basis in ${\mathcal{H}}$.
Under the WWM correspondence  the matrix product turns into
the Moyal product, or $\ast$-product, as follows:
\begin{eqnarray}
{\mbox{\boldmath{$U$}}} {\mbox{\boldmath{$V$}}} \longrightarrow
{\mathcal{U}}( x, y) & \ast & {\mathcal{V}}( x,y)  \equiv
\nonumber \\
& \equiv &
\exp\left[ i {\hbar\over 2}\left( {\partial^2\over \partial x^\prime_i
\partial y^i } -{\partial^2\over \partial x^j\partial y^\prime_j}
\right)
\right]
{\mathcal{U}}( x,y) {\mathcal{V}}( x^\prime ,y^\prime)
\vert_{\stackrel{\scriptstyle{}x^\prime=x}{\scriptstyle{}y^\prime=y}}
\nonumber\\
&\equiv &
\exp\left[ i {\hbar\over 2}\omega^{ab}
{\partial^2\over \partial \sigma^a \partial \xi^b }
\right]{\mathcal{U}}(\sigma) {\mathcal{V}}(\xi)
\vert_{\sigma=\xi},
\nonumber
\end{eqnarray}
where $\omega^{ab}$ is the symplectic form over the $2D$-dimensional
manifold with $\sigma \equiv (x,y)$ as coordinates.
By means
of the non-commutative $\ast$-product it is possible
to express the commutator
between two matrices, ${\mbox{\boldmath{$U$}}}$,
${\mbox{\boldmath{$V$}}}$, as
the \textit{Moyal Bracket\/} between their
corresponding Weyl symbols, ${\mathcal{U}}(x,y)$, ${\mathcal{V}}(x,y)$:
\[
{1\over \hbar}  \left[ {\mbox{\boldmath{$U$}}},
{\mbox{\boldmath{$V$}}}\right] \longrightarrow
\left\{ {\mathcal{U}}, {\mathcal{V}}\right\}_{\mathrm{MB}}
\equiv {1\over \hbar}\left(
{\mathcal{U}} \ast {\mathcal{V}} - {\mathcal{V}} \ast {\mathcal{U}}
\right)
\equiv \omega^{ij} \partial_i {\mathcal{U}}\circ \partial_j
{\mathcal{V}},
\]
where we introduced the $\circ$-product, which
corresponds to the ``even'' part of the $\ast$-product
\cite{strach}.  In the limit of vanishing deformation parameter
the Moyal bracket reproduces the Poisson bracket
\[
\lim_{\hbar\to 0}\left\{ {\mathcal{U}},
{\mathcal{V}}\right\}_{\mathrm{MB}}=
\left\{ {\mathcal{U}}, {\mathcal{V}}\right\}_{\mathrm{PB}}.
\]

The last step in the mapping of matrix theory
into a ``field model'' is carried
out through the identification of the
``\textit{deformation parameter\/}'' $\hbar$
with the inverse of $N$:
\[
\hbar\longrightarrow {2\pi\over N}\; : \quad \lim_{N\to\infty} f( N
)\longrightarrow\lim_{\hbar\to 0} f( \hbar ).
\]
Consistently, the large-$N$ limit of the
$\mathrm{SU}( N )$ matrix theory,
where the ${\mbox{\boldmath{$A$}}}_\mu$ quantum fluctuations freeze,
corresponds
to the quantum mechanical classical limit, $\hbar\to 0$, of the
WWM corresponding field theory\footnote{It is important to remark that
we are not attempting a WWM quantization of
the classical quenched field theory,
but we are only investigating the effects of
deforming the Lie algebraic structure.}
(from now on, we shall refer to the ``classical
limit'' without distinguishing between large-$N$ and small $\hbar$).\\
Finally, we can rewrite the trace operation as an integration over
$2D$ coordinates
\[
{(2\pi)^4\over N^3}
\mathrm{Tr}_{{\mathcal{H}}} \longrightarrow
\int \d^Dx  \d^Dy \equiv \int \d^{2D}\sigma
\]
and \cite{strach}
\[
{\mathcal{F}}_{\mu\nu}\equiv
\left\{ {\mathcal{A}}_\mu, {\mathcal{A}}_\nu\right\}_{\mathrm{MB}}
=\omega^{ab} \partial_a {\mathcal{A}}_\mu \circ  \partial_b
{\mathcal{A}}_\nu .
\]

After this technical detour, let us come back to the two terms
in the  reduced action (\ref{redym}), which are mapped by the WWM
correspondence into
\begin{eqnarray}
W_{\mathrm{qYM}}^{\mathrm{red.}}
&=&
-{1\over 4}\left({2\pi\over a}\right)^4
{N^4\over (2\pi)^4}\left( {2\pi\over N }\right)^2
{1\over g^2_{\mathrm{YM}}} \int \d^{2D}\sigma {\mathcal{F}}_{\mu\nu}
\ast {\mathcal{F}}^{\mu\nu}
\nonumber\\
&=&-{1\over 16}\left({2\pi\over a}\right)^4
{\left({N\over 2\pi}\right)}^2 {1\over g^2_{\mathrm{YM}}}
\nonumber \\
& & \qquad \times
\int \d^{2D} \sigma\omega^{ab}\partial_{[ a}
{\mathcal{A}}_{\mu} \circ \partial_{b ]} {\mathcal{A}}_{\nu}
\ast
\partial_{[ m} {\mathcal{A}}^{\mu}
\circ \partial_{n ]} {\mathcal{A}}^{\nu}
\omega^{mn},
\label{wym}
\\
W_{\mathrm{q\theta}}^{\mathrm{red.}}&=&-
{\theta g^2_{\mathrm{YM}}\over 16\pi^2}
\left({2\pi\over a}\right)^4
{N^4\over (2\pi)^4}\left( {2\pi\over N }\right)^2
\epsilon^{\mu\nu\rho\sigma}
 \int \d^{2D}\sigma {\mathcal{F}}_{\mu\nu}
\ast {\mathcal{F}}_{\rho\sigma}
\nonumber\\
&=&-{ \theta g^2_{\mathrm{YM}}  \over 64\pi^2}
\left({2\pi\over a}\right)^4
{\left({N\over 2\pi}\right)}^2 \epsilon^{\mu\nu\rho\sigma}
\nonumber \\
& & \qquad \times
\int \d^{2D} \sigma \omega^{ab}\partial_{[ a}
{\mathcal{A}}_{\mu} \circ \partial_{b]} {\mathcal{A}}_{\nu}
\ast
\partial_{[ m}{\mathcal{A}}_{\rho}\circ
\partial_{n ]} {\mathcal{A}}_{\sigma}
\omega^{mn}.
\label{wtheta}
\end{eqnarray}
We could also have written equation (\ref{wym})
without the $\ast$-product
between the two Moyal brackets due to the following  property of the
integration
over phase space in the absence of a boundary \cite{fairlie2},
$
\int \d^4\sigma {\mathcal{U}} \ast {\mathcal{V}}\equiv
\int \d^4\sigma {\mathcal{U}}  {\mathcal{V}}
$,
but extra terms will appear whenever boundaries are present \cite{cc}.
Since these extra terms are the ones in which we will be interested in,
we shall keep the $\ast$-product
in (\ref{wym}).\\
Before discussing the case $D=2$,
it can be useful to show how the classical
limit of $S_{\mathrm{q}}^{\mathrm{red.}}$
with $D=1$ is related to the action
of a bosonic string, which for simplicity we assume to be closed.
In this case only
the term $W_{\mathrm{qYM}}^{\mathrm{red.}}$ has to be taken into
account and gives
\begin{eqnarray}
W_{\mathrm{qYM}}^{\mathrm{red.}}{}_{N >> 1}&\approx &-{1\over 16}
\left( {2\pi\over a}\right)^4
{N^4\over (2\pi)^4} \left({ 2\pi\over N }\right)^2
{1\over g^2_{\mathrm{YM}}}
\nonumber \\
& & \qquad \times
\int \d^2 \sigma
\omega^{ab} \partial_{[ a}{\mathcal{A}}_{\mu} \partial_{b]}
{\mathcal{A}}_{\nu}   \omega^{mn}
\partial_{[ m}{\mathcal{A}}^{\mu} \partial_{n]}
{\mathcal{A}}^{\nu}
\nonumber\\
&=&-{1\over 16}\left( {2\pi\over a}\right)^4
\!\!
{\left({N\over 2\pi  }\right)}^2
\!\!
{1\over  g^2_{\mathrm{YM}} }  \int \d^2 \sigma
\left\{  {\mathcal{A}}_{\mu},
{\mathcal{A}}_{\nu} \right\}_{\mathrm{PB}}
\left\{{\mathcal{A}}^{\mu}, {\mathcal{A}}^{\nu} \right\}_{\mathrm{PB}},
\label{schildstr}
\end{eqnarray}
where we took into account that both the $\ast$ and $\circ$ products
collapse into the ordinary product in
the classical limit \cite{strach}.
Provided one appropriately rescales the gauge field, (\ref{schildstr})
reproduces the Schild action for the relativistic, bosonic string
\cite{schild,schildnoi}.
Let us remark that the Schild action is not
invariant under reparametrization
but only under the more restricted group of area
preserving diffeomorphisms;
this result establishes the known correspondence between
$\mathrm{SU}(\infty)$
and $\mathrm{sdiff}\left(\Sigma\right)$ \cite{bars}.
\begin{table}[ht]
\caption{\small{}This table summarizes the various
actions for the bulk $3$-brane
and boundary $2$-brane we used in the paper.}
\begin{tabular}{@{\extracolsep{\fill}}p{7.9cm}p{1.3cm}p{3.6cm}}
\hline
\centerline{}
\centerline{$p$\textbf{-brane Actions}}
&
\centerline{\textbf{Brane}} \centerline{\textbf{Fields}}
&
{\raggedright {$\,$} \\ \textbf{Symmetry}}
\\
\hline \hline
\centerline{$S _{\mathrm{DT}}\!\propto\!\int \d^4\sigma \sqrt{h}
h^{am} h^{bn}
\partial_{[ a} X^\mu \partial_{b]} X^\nu
\partial_{[ m} X_\mu \partial_{n]} X_\nu$}
&
\centerline{$X ^{\mu}(\sigma)$}
\centerline{$h _{mn} (\sigma)$}
&
{\raggedright
Reparametrization $+$\\
Conformal Invariance
}
\\
\hline
\centerline{$S _{\mathrm{Schild}}\!\propto\!\int \d^4\sigma
\left\{ X^\mu,  X^\nu \right\}_{\mathrm{PB}}
\left\{ X_\mu,  X_\nu \right\}_{\mathrm{PB}}$}
&
\centerline{$X ^{\mu}(\sigma)$}
&
{\raggedright
Volume Preserving\\
Diffeomorphisms
}
\\
\hline
\centerline{$S _{\mathrm{NG}}\!\propto\!\int \d^4\sigma
\sqrt{ -\mathrm{det}\left(  \partial_ m X_\mu \partial_n
X^\mu \right)  }$}
&
\centerline{$X ^{\mu}(\sigma)$}
&
{\raggedright
Reparametrization\\
Invariance
}
\\
\hline
\centerline{$S _{\mathrm{CS}}\!\propto\!
\epsilon_{\lambda\mu\nu\rho} \int \d^3\xi
X^{\lambda} \left\{   X^\mu,  X^\nu ,
X^{\rho} \right\}_{\mathrm{NPB}}$}
&
\centerline{$X ^{\mu}(\xi)$}
&
{\raggedright
Reparametrization\\
Invariance}
\\
\hline\hline\hline
\end{tabular}
\end{table}

Moving to the case $D=2$, it can be useful to recall that
the action for a $p$-brane can be written in several  different forms
\cite{neeman}. With hindsight, we need to recall
the conformally invariant $4$-dimensional  $\sigma$-model  action
introduced in \cite{dt}
\begin{equation}
S_{\mathrm{DT}}=-{\mu^4_0\over 4}
\int \d^4\sigma \sqrt{h}  h^{am} h^{bn}
\partial_{[ a} X^\mu \partial_{b]} X^\nu
\partial_{[ m} X_\mu \partial_{n]} X_\nu
\label{dtact}
\end{equation}
and the Chern--Simons membrane \cite{zaikov}
\begin{eqnarray}
S_{\mathrm{CS}}
&=&
-{\kappa\over 3!\times 4! } \epsilon_{\lambda\mu\nu\rho}
\int \d^3\xi
X^{\lambda} \epsilon^{abc} \partial_{ a} X^{\mu} \partial_b
X^\nu \partial_{c} X^{\rho} \nonumber\\
&=&-{\kappa\over 4! }   \epsilon_{\lambda\mu\nu\rho} \int \d^3\xi
X^{\lambda} \left\{   X^\mu,  X^\nu , X^{\rho} \right\}_{\mathrm{NPB}}.
\label{chs}
\end{eqnarray}
In equation (\ref{dtact}) the indices inside square brackets are
anti-symmetrized and
target  spacetime indices are saturated by
a flat metric $\eta_{\mu\nu}$.
In this $\sigma$-model approach $X^\mu$ would
be the coordinates of a $3$-brane
in $4$-dimensional target spacetime and the $\sigma^m$ are coordinates
on the $4$-dimensional world-volume. Moreover, $\eta_{\mu\rho}$ is the
flat Minkowski metric tensor in target spacetime,
while $h_{mn}(\sigma)$
is an independent, auxiliary, world-volume metric and provides
the reparametrization invariance of the model.
It can be worth to remind that,
once the auxiliary metric is algebraically
solved in terms of the induced
metric, i.e. $h_{mn}\propto \partial_m X \cdot \partial_n X$, then
$S_{\mathrm{DT}}$ turns into a Nambu--Goto type action. But,
the  Nambu--Goto  action
for a $3$-brane embedded in a $4$-dimensional
target spacetime is nothing
but the world-volume of the brane itself. Accordingly,
the constant $\mu^4_0$ in front of it can be identified with the
(constant) pressure inside the bag. Despite its non-trivial look,
$S_{\mathrm{DT}}$ does not describe transverse
propagating degrees of freedom, but
only a constant energy density and constant pressure,
non-dynamical spacetime domain.
All the dynamics is carried by the boundary of the domain,
in a way which
seems to satisfy the \textit{holographic principle\/}
\cite{holo} in a very
strict sense: all the non-trivial dynamical degrees of
freedom are confined to the
membrane enclosing the bag.
Among various kind of relativistic membranes the
Chern--Simons one is a very interesting object \cite{zaikov}.
In the action $S_{\mathrm{CS}}$ the $3$-volume
element of the membrane is represented
by the Nambu--Poisson brackets, while $\kappa$ is a constant
with dimension of energy per unit $3$-volume. The presence of the
Nambu--Poisson brackets suggests a new kind of formulation of both
classical and quantum mechanics for such an object,
which is worth investigating by itself \cite{flato}.\\
Moreover, the formal structures of (\ref{wym}, \ref{wtheta}) and
(\ref{dtact}, \ref{chs})
are so similar that one expects some kind of relationship among these
actions. On the other hand, we notice that while the
$W_{\mathrm{qYM}}^{\mathrm{red.}}$  action
is defined over a \textit{flat\/} phase space,
$S_{\mathrm{DT}}$ involves integration
over a \textit{curved\/} world-volume.
In the latter case a Moyal deformation
would be no longer valid, due to the lack of  associativity of the
$\ast$-product,
and a Fedosov deformation quantization would be required
\cite{cc2}. However, the Weyl symmetry of the
$S_{\mathrm{DT}}$ also suggests to
restrict\footnote{In two dimensions every metric is
conformally flat. In
our case, i.e. four dimensions, we need to require conformal flatness.}
the world metric to the conformally flat sector:
\begin{equation}
h_{mn}= e^{2\phi(\sigma)}  \eta_{mn}.
\label{gauge}
\end{equation}
We shall not consider in this paper the quantum conformal
anomaly, which could spoil the choice (\ref{gauge}),
and concentrate our attention only over the
equivalence between classical actions.
Furthermore, we can implement  the following
relation between the Poisson bracket
and the simplectic form  in four dimensions:
\[
\partial_{[a } X^\mu   \partial_{b] } X^\nu= {1\over 4}\omega_{ab}
\left\{   X^\mu, X^\nu \right\}_{\mathrm{PB}}.
\]
Thus, we find that the $S_{\mathrm{DT}}$ action in a conformally flat
background geometry looks like a generalized
Schild action \cite{schild}
for a $3$-brane:
\begin{equation}
S_{\mathrm{DT}}=-{\mu^4_0\over 16} \int \d^4\sigma
\left\{ X^\mu,  X^\nu \right\}_{\mathrm{PB}}
\left\{ X_\mu,  X_\nu \right\}_{\mathrm{PB}}
\equiv S_{\mathrm{Schild}}.
\label{sch}
\end{equation}
The alleged correspondence between $W_{\mathrm{qYM}}^{\mathrm{red.}}$
and
$S_{\mathrm{DT}}$ can be seen as follows.
By choosing $D=2$ in (\ref{wym}), we find
\begin{eqnarray}
W_{\mathrm{qYM}}^{\mathrm{red.}}{}_{N>>1}&\approx &-{1\over 16}
\left( {2\pi\over a}\right)^4
 \left( { N\over 2\pi   } \right)^2
{1\over  g^2_{\mathrm{YM}}}
\nonumber \\
& & \qquad \times
\int \d^4 \sigma
\omega^{ab} \partial_{[ a}
{\mathcal{A}}_{\mu} \partial_{b]}  {\mathcal{A}}_{\nu}  \omega^{mn}
\partial_{[ m}{\mathcal{A}}^{\mu}\partial_{n]}
{\mathcal{A}}^{\nu}
\nonumber\\
&=&-{1\over 16}\left( {2\pi\over a}\right)^4
\!\!
\left( { N\over 2\pi   } \right)^2
\!\!
{1\over  g^2_{\mathrm{YM}}}
\int \d^4 \sigma \left\{  {\mathcal{A}}_{\mu},
{\mathcal{A}}_{\nu} \right\}_{\mathrm{PB}}
\left\{{\mathcal{A}}^{\mu}, {\mathcal{A}}^{\nu} \right\}_{\mathrm{PB}};
\nonumber
\\
W_{\mathrm{q\theta}}^{\mathrm{red.}}&=&-
{\theta g^2_{\mathrm{YM}}\over 16\pi^2}
\left({2\pi\over a}\right)^4
{N^4\over (2\pi)^4}\left( {2\pi\over N }\right)^2
\epsilon^{\mu\nu\rho\sigma}
\int_\Sigma \d^{4}\sigma {\mathcal{F}}_{\mu\nu}
{\mathcal{F}}_{\rho\sigma}\nonumber\\
&=&-{ \theta g^2_{\mathrm{YM}}  \over 64\pi^2}
\left({2\pi\over a}\right)^4
{\left({N\over 2\pi}\right)} \epsilon^{\mu\nu\rho\sigma}
\nonumber \\
& & \qquad \times
\int_\Sigma \d^{4} \sigma \omega^{[ ab} \partial_{a}
{\mathcal{A}}_{\mu} \partial_{b} {\mathcal{A}}_{\nu}
\partial_{m} {\mathcal{A}}_{\rho} \partial_{n} {\mathcal{A}}_{\sigma}
\omega^{mn]}\nonumber\\
&=&-{ \theta g^2_{\mathrm{YM}}  \over 32\pi}
\left({2\pi\over a}\right)^4
{\left({N\over 2\pi}\right)}^2 \epsilon^{\mu\nu\rho\sigma}
\int_{\partial\Sigma }\d^3 s {\mathcal{A}}_{\mu}
\left\{ {\mathcal{A}}_{\nu}, {\mathcal{A}}_{\rho},
{\mathcal{A}}_{\sigma}\right\}_{\mathrm{NPB}}.
\nonumber
\end{eqnarray}
By rescaling the fields according with\footnote{It can
be useful to list the  dimensions of
various quantities in  natural units. The main reason is that
quenched, dimensional reduced,
gauge variables have not canonical dimensions:
\begin{eqnarray}
&&
\left[{\mbox{\boldmath{$D$}}}_\mu\right]
=
\left[{\mathcal{A} }_\mu\right]
=
\mathrm{length},
\qquad \left[{\mathcal{F} }_{\mu\nu}\right]= (\mathrm{length})^2,
\nonumber\\
&& \left[X^\mu\right]= \mathrm{length},\nonumber\\
&& \left[a\right]= \mathrm{length},\qquad   \left[\Lambda\right]=
(\mathrm{length})^{-1}, \nonumber\\
&& \left[\sigma^m\right]=1 ,\qquad \left[ \left\{ X^\mu,  X^\nu
\right\}_{\mathrm{PB}}\right]=(\mathrm{length})^2,\nonumber\\
&& \left[\mu_0\right]= (\mathrm{length})^{-1},\qquad
\left[\kappa\right]= (\mathrm{length})^{-4}.
\nonumber
\end{eqnarray}
}
\begin{eqnarray}
{\mathcal{A}}_\mu  &\longrightarrow&
\left({2\pi\over N}\right)^{1/4} X_\mu
\nonumber \\
{\mathcal{F}}_{\mu\nu}
&\longrightarrow&
\left({2\pi\over N} \right)^{1/2}
 \left\{ X_\mu,  X_\nu \right\}_{\mathrm{PB}}
\nonumber
\end{eqnarray}
we get
\begin{eqnarray}
W_{\mathrm{qYM}}^{\mathrm{red.}}
+
W_{\mathrm{q}\theta}^{\mathrm{red.}}
&=&
-{1\over 16}\left( {2\pi\over a}\right)^4
{1\over  g^2_{\mathrm{YM}}}
\int \d^4 \sigma \left\{  {X}_{\mu},  { X}_{\nu}
\right\}_{\mathrm{PB}}
\left\{{ X}^{\mu}, { X}^{\nu} \right\}_{\mathrm{PB}}\nonumber\\
& & \quad -
{\theta g^2_{\mathrm{YM}}  \over 32\pi}
\left({2\pi\over a}\right)^4 \epsilon^{\mu\nu\rho\sigma}
\int_{\partial\Sigma }\d^3 s  { X}_{\mu}
\left\{ { X}_{\nu}, { X}_{\rho}, { X}_{\sigma}\right\}_{\mathrm{NPB}}
,
\nonumber
\end{eqnarray}
which matches the sum of the actions
$S_{\mathrm{DT}}$ and $S_{\mathrm{CS}}$
provided we identify
\begin{equation}
\mu_0^4
\longleftrightarrow
{1\over 4\pi }\left( {2\pi\over a}\right)^4
{4\pi\over g^2_{\mathrm{YM}}}
\label{bag}
\end{equation}
and
\[
\kappa
\longleftrightarrow
{\theta g^2_{\mathrm{YM}}  \over 32\pi}\left({2\pi\over
a}\right)^4 .
\]
As a consistency check of our
\textit{dynamically generated bag pressure\/}
consider equation (\ref{bag}) in the strong coupling regime,
where it is conventionally assumed
$g^2_{\mathrm{YM}}/4\pi \simeq 0.18$.
If we identify the inverse lattice spacing $2\pi/a$ with the QCD
scale
$\Lambda_{\mathrm{QCD}}\simeq 200 \, \mathrm{MeV}$,
then equation (\ref{bag}) provides
\[
\mu_0^4\equiv  {1\over 4\pi  }\left( 200 \, \mathrm{MeV} \right)^4
{1\over 0.18}.
\]
The actual value of $\mu_0$ is close to $\Lambda_{\mathrm{QCD}}$;
this is not a bad result, compared with the phenomenological
value $\mu_0\simeq 110 \, \mathrm{MeV}$,
if one takes into account the uncertainty
on the  value of $\Lambda_{\mathrm{QCD}}$, i.e.
$120 \,\mathrm{MeV}\le \Lambda_{\mathrm{QCD}}\le 350\,\mathrm{MeV}$.

\begin{figure}
\caption{\small{}The figure shows the web of
relationships among the various actions
discussed in this note. It can be read as
``map'' to move from the Yang--Mills
action, in the lower left corner, to the complete
bag action in the upper right corner.}
\begin{center}
\includegraphics{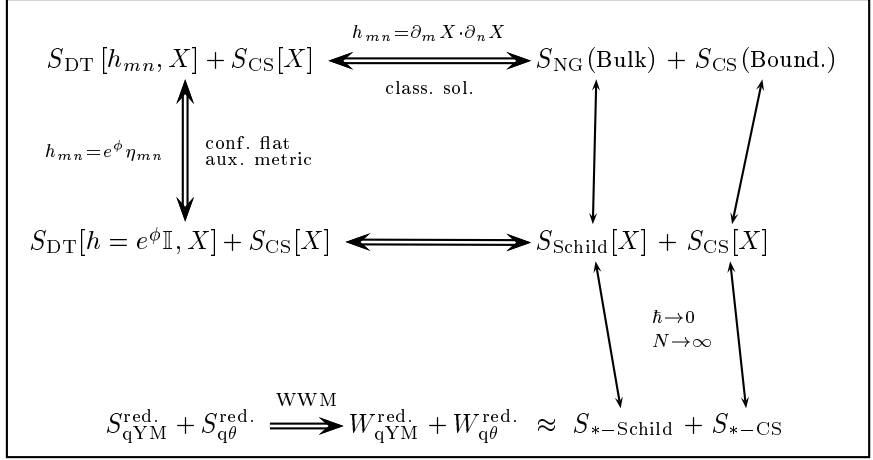}
\end{center}
\end{figure}

In summary, we have shown that reduced,
quenched $\mathrm{SU}( N)$  gauge theory
can fit, in $4$-dimensions and in the large-$N$ limit,
not only strings
but $3$-branes with a dynamical boundary as well.
The WWM correspondence maps the original matrix
action (\ref{redym})
into the phase space action (\ref{wym}, \ref{wtheta}),
where the gauge field strength is replaced by a
Moyal bracket for the Weyl
symbol of the matrix gauge field.
The  large-$N$, or classical, limit of (\ref{wym}) reproduces
the Dolan--Tchrakian action $S_{\mathrm{DT}}$ for a $3$-brane in the
conformally flat background geometry (\ref{gauge}) and a Chern--Simons
action for its boundary.
In even dimensional target spacetime the  $S_{\mathrm{DT}}$
functional matches the Schild action (\ref{sch}). In analogy with
the string case, the action (\ref{sch}) is not invariant under
reparametrization but it is only invariant under residual world-volume
preserving diffeomorphisms. With hindsight, this is not a surprise:
the reduced symmetry seems to be the memory
of the constant inverse volume
factor  $( 2 \pi / a ) ^{4}$
in front of the original reduced quenched action (\ref{redym}).
A similar conclusion was obtained in \cite{sez}, where the light-cone
gauge choice leads to a residual
$p$-volume preserving diffeomorphisms invariance,
while we have $(p+1)$-world-volume preserving diffeomorphisms.
In the limiting case $p=3$, $D=4$ the $S_{\mathrm{DT}}$
action degenerates into a pure
volume term with no proper dynamics.
All the physical degrees of freedom are
carried by the Chern--Simons membrane enclosing the bag.
By tracing back the
bulk and boundary terms to the original Yang--Mills
action the following
correspondence will show up:
\begin{eqnarray}
\hbox{``glue''}&:& \quad S_{\mathrm{qYM}}^{\mathrm{red.}}
\longleftrightarrow
S_{\mathrm{DT}}\propto \hbox{ Bulk Volume}\nonumber\\
\hbox{ instantons}&:& \quad S_{\mathrm{q}\theta} ^{\mathrm{red.}}
\longleftrightarrow S_{CS}\propto \hbox{Boundary Membrane} .
\nonumber
\end{eqnarray}
Finally, an order of magnitude estimate of the induced
bag constant results to
be in agreement with the phenomenological value,
and suggests a model for
hadrons as QCD vacuum bubbles bounded by Chern--Simons membranes.
This new formulation of the hadronic bag model and its
generalization  to the case of higher dimensional
branes are currently under investigation \cite{cc3}.

\end{document}